\documentclass[aps,prl,twocolumn,showpacs,nofootinbib]{revtex4}
\usepackage{amsmath,amssymb,graphicx}
\usepackage{epsf}
\usepackage{bm}

\newcommand{\nc}{\newcommand}
\nc{\postscript}[2]
{\setlength{\epsfxsize}{#2\hsize}\centerline{\epsfbox{#1}}}
\nc{\non}{\nonumber}
\nc{\hc}{\hbox {h.c.}} \nc{\re}{\hbox {Re}} 
\nc{\mev}{\hbox {MeV}} \nc{\gev}{\;\hbox {GeV}} \nc{\tev}{\;\hbox {TeV}}
\def\lsim{\mathrel{\raise.3ex\hbox{$<$\kern-.75em\lower1ex\hbox{$\sim$}}}}
\def\gsim{\mathrel{\raise.3ex\hbox{$>$\kern-.75em\lower1ex\hbox{$\sim$}}}}

\nc{\etal}{{\it et al.}}
\nc{\Lsp}{\;\;\;\;\;\;\;\;\;\;}  \nc{\LLLsp}{\lspace \lspace}
\nc{\lsp}{\;\;\;\;\;\;}
\nc{\spac}{\;\;\;}
\nc{\noi}{\noindent}
\nc{\beq}{\begin{equation}}   \nc{\eeq}{\end{equation}}
\nc{\bea}{\begin{eqnarray}}   \nc{\eea}{\end{eqnarray}}
\nc{\baa}{\begin{array}}      \nc{\eaa}{\end{array}}
\nc{\bit}{\begin{itemize}}    \nc{\eit}{\end{itemize}}
\nc{\ben}{\begin{enumerate}}  \nc{\een}{\end{enumerate}}
\nc{\bce}{\begin{center}}     \nc{\ece}{\end{center}}

\def\calQ{{\cal Q}}
\def\calU{{\cal U}}

\def\sq2{\sqrt{2}}

\def\ph{\varphi}

\def\m4{m^4(\ph)}
\def\mn2{m_n^2}

\def\v5{V^{(5)}}

\begin{document}

\title{\begin{flushright}
       \mbox{\normalsize \rm UMD-PP-08-22}
       \end{flushright}
       \vskip 20pt
Radion Mediated Flavor Changing Neutral Currents}
\author{Aleksandr Azatov}
\author{Manuel Toharia}
\author{Lijun Zhu}
\affiliation{
Maryland Center for Fundamental Physics, Department of Physics, University of Maryland\\
College Park, MD 20742, USA
}

\date{\today}

\begin{abstract}
In the context of a warped extra-dimension with Standard Model fields in the
bulk, we obtain the general flavor structure of the Radion couplings to
fermions and show that the result is independent on the particular
nature of the Higgs mechanism (bulk or brane localized). These
couplings will be generically misaligned with respect to the fermion
mass matrix when the fermion bulk mass parameters are
not all degenerate. When the Radion is light enough, the generic size
of these tree-level flavor changing couplings will be strongly
constrained by the experimental bounds on $\Delta F=2$ processes. At
the LHC the possibility of a heavier Radion decaying into top and
charm quarks is then considered as a promising signal to probe
the flavor structure of both the Radion sector and the whole scenario.

\end{abstract}

\maketitle 

Introducing a warped extra-dimension in such a way as to create an
exponential scale hierarchy between the two boundaries of the extra
dimension~\cite{RS1} has generated a lot of attention in the recent
years as a novel approach to solve the hierarchy problem. 
By placing the Standard Model (SM) fermions in the bulk of the extra
dimension it was then realized that one can simultaneously address
the fermion mass hierarchy puzzle~\cite{a}.
In this context the main constraints come from
precision electroweak bounds as well as from low energy flavor
violating processes
\cite{RSeff, Burdman, Agashe, Weiler}, 
pushing the scale of new physics  (the mass of the lowest KK
excitations) to several TeV. 
In these scenarios, the metric fluctuations contain a scalar degree of freedom - the Radion, whose mass and couplings
could make it the first {\it new physics} state to be discovered at
the LHC. In the original RS1 setup~\cite{RS1}, the Radion
phenomenology was extensively studied and analyzed including the
possibility of some amount of mixing with the Higgs
scalar~\cite{GRW,CGK,Korean}. But it wasn't until
relatively recently~\cite{rizzo,CHL,mtgamma} that Radion interactions with
bulk SM fields were fully considered. In this letter we want to extend these
last investigations to include the full fermion flavor structure to the
Radion couplings and show that as opposed to the original RS1
scenario, there is a prediction for generic flavor violating Radion couplings
to fermions.
The spacetime we consider takes the usual Randall-Sundrum form~\cite{RS1}:
\bea
ds^2
= \frac{R^2}{z^2}\! \Big(\eta_{\mu\nu} dx^\mu dx^\nu -dz^2\Big),
\label{RS}
\eea
with the UV (IR) branes localized at $z = R$ ($z =
R^\prime$).
The Radion can be parametrized by the following scalar perturbation of metric:
\bea
\hspace{-1cm} ds^2  \label{metricpert}
&=& \left(\frac{R}{z} \right)^2(e^{-2F}\eta_{\mu\nu} dx^\mu dx^\nu - (1+2F)^2 dz^2)
\eea
Demanding that the perturbed metric solves the Einstein equation and
that the Radion field is canonically normalized, we get
\bea
F
\ =\ \frac{r(x)}{\Lambda_r} \frac{z^2}{R'^2}
\eea
where $r(x)$ is the corresponding canonically normalized Radion graviscalar with its
associated interaction scale $\Lambda_r=\sqrt{6}\frac{\!R}{R'}M_{Pl}$.
We assume that some
unknown dynamics (e.g. the Goldberger-Wise
mechanism~\cite{Goldberger:1999uk}) will fix the inter-brane distance and give a
positive mass squared to the Radion, and that it gives negligible back-reaction to the metric. 
The couplings between bulk SM fermions and the Radion are calculated
in~\cite{CHL} in the case of one generation, with a brane localized
Higgs. We are interested here in the flavor structure of these
couplings when all families of fermions are considered, and
for the more general case of a 5D bulk Higgs  $\cal{H}$ \cite{bulkhiggs}. 
To this end let us focus on the up-sector of the simple setup
in which we consider the 5D fermions $\calQ_i$, $\calU_i$, with flavor indices
$i,j=1,2,3$. They contain the 4D SM $SU(2)_L$ doublet and singlet fermions
respectively with a 5D action   
\bea\label{fermionaction}
&&\hspace{-.5cm} S_{\text{fermion}}\!=\!\int d^4x dz \sqrt{g} \Big[
{i \over 2} \left(\bar{\calQ_i} \Gamma^A {\cal D}_A \calQ_i -
{\cal D}_A \bar{ \calQ_i} \Gamma^A \calQ_i\right)  \non\\
&&\hspace{-.5cm} + {c_{q_i} \over R} \bar{ \calQ}_i {\calQ}_i +
({\calQ} \rightarrow {\calU}) +\left(Y_{ij}\sqrt{R}\ \bar{\calQ_i}
{\cal H} \calU_j + h.c.\right)
\Big]
\eea
where $\frac{c_{q_i}}{R}$, $\frac{c_{u_i}}{R}$ are the 5D fermion
masses, and we choose to work in the basis where they are diagonal in 5D
flavor space. 
The bulk Higgs acquires a nontrivial VEV $v(z)$ localized towards the IR
brane solving the Planck-weak hierarchy problem. Note that the special case
of brane-localized Higgs case can be treated as a limit of the bulk Higgs case.
After writing the 5D fermions in two component notation, 
$\calQ_i=\left(\begin{array}{c}\calQ^i_L\\ \bar{\calQ}^i_R\end{array}\right)$
and
$\calU_i=\left(\begin{array}{c}\calU^i_L\\ \bar{\calU}^i_R\end{array}\right)$,
we perform a ``mixed'' KK decomposition as
\bea
\calQ^i_L(x,z) & =&   Q_L^{ij}(z)\, q_L^j (x)  + ...\label{kkdecomp1}\\
\bar{\calQ}^i_R(x,z)& =& Q_R^{ij}(z) \bar{u}^j_R(x)+...\label{kkdecomp2}\\
\calU^i_L(x,z) & =&   U_L^{ij}(z)  q_L^j(x) +...\label{kkdecomp3}\\
\bar{\calU}^i_R(x,z)& =&  U^{ij}_R (z)\, \bar{u}_R^j (x) +...\label{kkdecomp4}
\eea
where we have only written the 4D SM fermions $q^j_L(x),\ u^j_R(x)$
and where $Q^{ij}_{L,R}(z),\ U^{ij}_{L,R}(z)$ are the corresponding profiles along the extra
dimension. The fields $q_L^i(x)$ and $u_R^j(x)$ verify the Dirac equation
\bea
-i \bar{\sigma}^{\mu} \partial_\mu q_L^{i} + m_{ij}\, \bar{u}_R^j &=& 0, \\
-i \sigma^{\mu} \partial_\mu \bar{u}_R^i + m_{ij}\, q_L^j& =& 0,
\eea
with the 4D SM fermion mass matrix $m_{ij}$ not necessarily
diagonal in flavor space. 
The couplings between Radion and SM fermions can be calculated by
inserting the perturbed metric of Eq.~(\ref{metricpert}) and the 5D fermion
KK decompositions of Eqs.~(\ref{kkdecomp1}-\ref{kkdecomp4}) into the action
of Eq.~(\ref{fermionaction}).
To proceed we used a perturbative approach treating the 4D
fermion masses $m_{ij}$ as small expansion parameters (i.e. we assumed
$m_{ij}R'\ll 1$) keeping only first order terms. In this limit, the profiles $Q^{ij}_L(z)$ and
$U^{ij}_R(z)$ match the simple wave-functions for massless zero-modes.
No other explicit profile solution is required since we just need to
properly insert and use the KK equations
for $Q^{ij}_R(z)$ and $U^{ij}_L(z)$ into Eq.~(\ref{fermionaction}).
A subtlety however is that the 5D bulk
Higgs field perturbation contains itself some Radion degree of
freedom. This can be seen from solving the Higgs equations of motion
in the perturbed background of Eq.~(\ref{metricpert}), which requires the KK
expansion of the 5D Higgs field to be of the form
\bea
{\cal{H}}(x,z) &=& v(z) - \frac{z^3 v^\prime(z)}{R'^2}
\left[1-\left(\frac{R'}{z}\right)^2\right]\ \frac{r(x)}{\Lambda_r}+\cdots\hspace{.8cm}
\eea
where 
the $\cdots$ contain the 4D light Higgs and the rest of the Higgs KK
modes. This result 
gives an additional contribution to the Radion coupling to
fermions. It is possible to show that the general formula for the
Radion coupling to SM fermions is
\bea
\label{Radioncoupling}
\hspace{-2cm}&&-\frac{r(x)}{\Lambda_r} \left(q_L^{i} u_R^{j} +
\bar{q}_L^{i} \bar{u}_R^{j} \right)  {m^u_{ij}}
\left[{\cal I}({c_{q_i}}) + {\cal I}({-c_{u_j}})\right]
\eea
where we have defined
\bea
{\cal I}(c)= \left[\frac{(\frac{1}{2}-c)}{1-{(R/R')}^{1-2c}}+c\right] \approx \Big\{ \begin{array}{c} c
  \, \,(\, c\, > \,1/2\,) \\ \frac{1}{2}\,\, (\, c \,<
  \,1/2\,) \end{array}.
\eea
For one generation of fermions, this result agrees with the
formulae obtained in \cite{CHL} and it can also be understood from the
following intuitive argument. When the 4D SM fermion mass is generated near the IR brane, its dependence on
$\frac{1}{R^\prime}$ is
\begin{equation}
m_{ij}\propto f(c_{q_i}) f(-c_{u_j})  \frac{R}{R^\prime}
\end{equation}
with $f(c)$ proportional to the zero mode wavefunction of the fermions evaluated at IR brane
\begin{eqnarray}
f(c) = \sqrt{\frac{1-2c}{1-(R/R^\prime)^{1-2c}}}
\end{eqnarray}
Since the Radion is basically a fluctuation of the IR brane location,
its couplings with SM fermions can also be obtained by
replacing $\frac{1}{R^\prime} \rightarrow
\frac{1}{R^\prime}(1-\frac{r}{\Lambda_r})$ in the fermion mass
matrix~\cite{CHL}.
Then it is easy to check that we reproduce
the result of Eq.~(\ref{Radioncoupling}). 
Non-univeralities in the term $\left[{\cal I}({c_{q_i}}) + {\cal
    I}({-c_{u_j}})\right]$ will lead to a misalignment between the
Radion couplings and the fermion mass
matrix\footnote{This will remain true in the presence of fermion brane
  kinetic mixings although the flavor structure of
  (Eq.\ref{Radioncoupling}) will be modified.}. 
After diagonalization of the
fermion mass matrix, flavor violating couplings will be generated and
can be parametrized as
\begin{equation}
\mathcal{L}_{FV} = \frac{r}{\Lambda_r}(\bar{u}^i_L u^j_R  a_{ij}\sqrt{m_i m_j}+
\text{h.c.})\qquad (i \ne j) \label{rfv}
\end{equation}
where $u^{i}$ are the quark mass eigenstates with masses $m_{i}$. 
The extension to the down quark sector and charged leptons is
immediate.

To study the consequences of this result, we will consider models with flavor
anarchy i.e. where all the hierarchies in the fermion sector are explained by the warp factors
and all 5D Lagrangian parameters are of the same order~\cite{Agashe}. 
In this class of models the natural size of $a_{ij}$ is 
\begin{equation}
a_{ij} \sim (\Delta {\cal I}_{ij})\sqrt{\frac{ f(c_{q i}) f(-c_{u j})}{f(c_{q j}) f(-c_{u i})}}
\end{equation}
where $\Delta{\cal I}_{ij} \sim O(0.1)$\footnote{This estimate is only valid for
  models that explain the Planck-weak hierarchy. But for little RS
  models~\cite{LRS}, the deviation could be a few times larger.} is the
deviation of $\left[{\cal I}({c_{q_i}}) + {\cal I}({-c_{u_j}})\right]$
from its mean value.
We perform a scan over the 5D fermion masses and ``anarchical'' Yukawa
couplings leading to the observed SM fermion masses and
CKM mixing angles and obtain a distribution for the parameters
$a_{ij}$. For example, the average values
of the parameter $a^d_{12}$ and $a^d_{21}$ are of order $\sim 0.07$
and $70\%$ of the time they are distributed between $0.03<a^d_{12}, \,
a^d_{21}<0.12$. The average value of the parameter
$a^u_{23}$($a^u_{32}$) are $\sim 0.08$($0.05$) and $70\%$ of the times
they are between $0.03<a^u_{23}<0.13$ ($0.01<a^u_{32}<0.09$).

The first thing to study is how constrained are the Radion
parameters due to low energy observables such as $\Delta F=2$
processes. The processes mediated by virtual Radion exchange will have
the following flavor structure
\begin{eqnarray}
\bar q^\alpha_{i L}q^\alpha _{j R}\bar q^\beta_{k L}q^\beta_{ n R}
\end{eqnarray}
$\alpha,\beta$ are color indices and $i,j,k,n$ are flavor
indices. One can see that this interaction can be parametrized by
standard $Q_2,Q_4$ operators (see for
example~\cite{Buras:1998raa}). The strongest constraints will come from
$\epsilon_K$; and the model independent constraint on the size
of new physics contributions to the imaginary part of the Wilson coefficient
$C_{4K}$, renormalized at the scale $50$ GeV, is $\text{Im} C_{4} \lesssim
1.2\times10^{-10} \text{TeV}
^{-2}$~\cite{Bona:2007vi}\footnote{We used the RG equations in
  \cite{Bagger:1997gg}. Constraints on the coefficient $C_2$
  are weaker by a factor of five and the bounds from $B_d$ mixing are
  weaker by an order of magnitude, so we ignored them in the present
  analysis.}.
From Eq.~(\ref{rfv}) it is easy to compute the contribution from a tree-level
Radion exchange as
$\text{Im}(C^{Radion}_{4})\approx m_d m_s \text{Im}(a^d_{12} a^{d\ast}_{21})/(\Lambda_r^2 m_r^2)$
and therefore the experimental bound requires that $a_{ds}/(\Lambda_r m_r)<0.44 \text{ TeV}^{-2}$, 
where we define $a_{ds} \equiv \sqrt{|a^d_{12} a^{d\ast}_{21}|}$ and assume order one phase. 
\begin{figure}[t]
\vspace{-.2cm}
\center
\includegraphics[width=8.7cm]{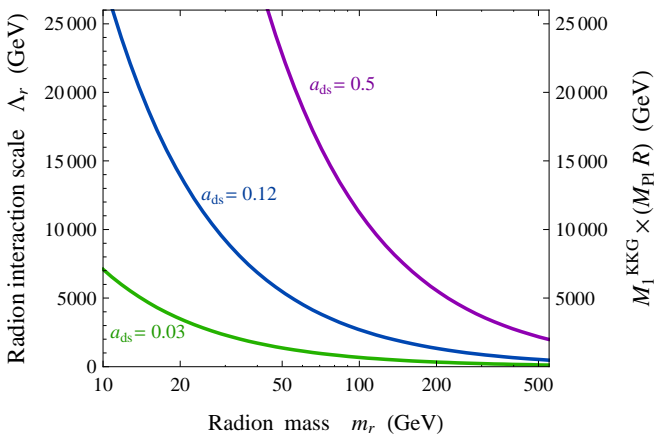}\ \ \ \ \
\vspace{-.4cm}
\caption{Bounds in the ($m_r-\Lambda_r$) plane coming from
  $\epsilon_K$ for different values of the flavor violating parameter
  ${a}_{ds}=\sqrt{|a^d_{12} a_{21}^{d*}|}$. In flavor anarchy
  models~\cite{Agashe}, typical values for  ${a}_{ds}$ range between
  $0.03$ and $0.12$. In Little RS~\cite{LRS} this  parameter can reach
  values a few times larger. One can relate the scale $\Lambda_r$ to
  the mass of the lightest KK gluon as $M^{KKG}_{1}\simeq\Lambda_r/(M_{Pl}R)$, as shown on the RHS of the figure.
}
\label{epsilonK}
\end{figure}
In Fig.~\ref{epsilonK}, we show the bounds for different
values of $a_{ds}$  
in the ($m_r$, $\Lambda_r$) plane. The scale $\Lambda_r$ is directly related to the
lightest KK gluon mass by $M^{KKG}_{1}\simeq\Lambda_r/(M_{Pl}R)$, and so 
one can easily convert bounds on the KK mass into bounds on $\Lambda_r$.
It is also interesting to note that the bounds from flavor physics give strong
constraints for a very light Radion, precisely the hardest possibility to probe at
the LHC due to its dominant hadronic decay channels. A light Radion
with flavor violating couplings can also become a top
quark decay product, in processes such as $t\to r c$ or $t\to r u$,
where $u$ and $c$ are the up and charm quarks. We have checked that, due
to the suppressed couplings coming from $\Lambda_r$, this signal~\cite{aguilar}
will not be visible at the LHC unless the flavor violating parameters
$a_{i3}$ or $a_{3i}$ take unnaturally large values (${\cal O}(1)$).

\begin{figure}[b]
\vspace{-.2cm}
\hspace{-.2cm}
\includegraphics[width=8.2cm]{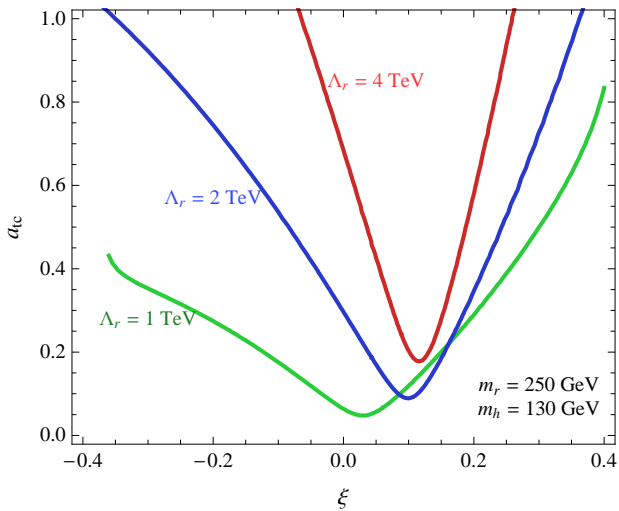}\ \ \ \ \
\vspace{-.4cm}
\caption{Contours in the ($\xi-a_{tc}$) plane of the estimated signal
  significance $S/\sqrt{B}=3$ for the process ($pp\to r\to tc$) at
  the LHC for $300$ fb$^{-1}$ of data. $\xi$ is the Higgs-Radion mixing
  parameter and $a_{tc}$ is the flavor violating parameter which gives rise to
  the Radion coupling to top-charm.
}
\label{reach plot}
\end{figure}

For a heavier Radion ($\gtrsim 200$ GeV), the most
promising discovery channel would be $r \rightarrow Z Z
\rightarrow 4l$ due to its clean signal. Translating the LHC Higgs Search analysis~\cite{CMS}
into Radion LHC reach, one finds that both CMS and ATLAS should
separately be able to claim discovery for $\Lambda_{r}\lesssim 5$ TeV with
30 fb$^{-1}$ of data~\cite{MTNO}. To study the flavor structure of
such a heavy Radion, we consider the channel $r \rightarrow \bar{t} c,
t \bar{c}$ (see for example \cite{ctburdman} in top-condensation models). We
define $a_{tc} \equiv \sqrt{(|a^u_{23}|^2+|a^u_{32}|^2)/2}$ to parametrize the
flavor violating coupling between Radion and top charm. 
The signal we focus on is $p\,p \rightarrow t c \rightarrow
b\, l\, \nu\, c$ (where $l$ stands for electrons and muons). 
And the main backgrounds are: (i)  $p,p
\rightarrow t\, j \rightarrow b\, l\,
\nu\,\, j$; (ii)  $p,p \rightarrow W\,j \, j \rightarrow l\, \nu\,
j\, j$, where one of the light jet is mistagged as b quark;
(iii)  $p\, p \rightarrow W\, \bar{b}\,b \rightarrow b\,
\bar{b}\, l\, \nu$, where one of the b jet is mistagged;  
(iv) $p\, p\rightarrow \bar{t} \, t \rightarrow b \, l^+ \, \nu
\, \bar{b} \, l^- \, \bar{\nu}$ where one b jet is mistagged and one
of the charged lepton is lost in beam pipe ($|y_l| > 2.5$) or it is
merged with one of the jets ($\Delta R_{jl} < 0.6$). 
We use CalcHEP~\cite{calchep} and PYTHIA 2.6~\cite{pythia}
to obtain both signal and background cross sections and estimate 
the potential LHC reach for this signal. For this we fix the Radion
interaction scale to $\Lambda_r=2$ TeV, and use three
different values for its mass, 
$m_r=250,300$ and $350$ GeV. 
We impose lepton and jet acceptance cuts on transverse
momenta  $p_T^{j,l}  > 20$ GeV,
on rapidities, $|y_{j,l}| < 2.5$, and on angular separation
$\Delta R_{lj} > 0.6$ and $\Delta R_{jj} > 0.6$.
We assume that neutrino momentum can be reconstructed.
We demand additionally that the
total event's invariant mass reconstructs to the
Radion's mass $M_{bl\nu j} \in (m_r-5\,$ GeV, $m_r+5\,$ GeV), and
that the $bl\nu$ invariant mass reconstructs to the top mass
$M_{bl\nu} \in (170\,$ GeV, $180\,$ GeV). We also tighten the rapidity
cut on the light jet, $|y_j| < 1.5$. 
We assume that the Radion would have been discovered through $r
\rightarrow Z Z$ channel and thus measured its mass $m_{r}$. Because
the Radion decay width is extremely small ($\Gamma_r<0.15$ GeV in this
mass range), the window to use for the total invariant mass is controled
by the experimental jet energy resolution (we used a
window of $\pm 5$ GeV).
The results are shown in Table.~\ref{tctable}.
As noted in~\cite{mtgamma}, a small amount of Higgs-Radion mixing~\cite{GRW},
parametrized by the Lagrangian parameter $\xi$, can dramatically
reduce the principal Radion decay channels. This could then enhance secondary decay
channels, such as $r\rightarrow \gamma \gamma$, and in this case
$r\rightarrow \bar{t}c (t\bar{c})$. In Fig.~\ref{reach plot}
we plot contours for the LHC reach in the $(a_{tc}$ vs. $\xi)$ plane, for $m_r = 250$ GeV
and different values of $\Lambda_r$. We can see that at least for some ranges
of $\xi$, the LHC should be able to probe typical values of $a_{tc}$
in flavor anarchy models. Of course a more realistic study of
this signal should be carried out, including a full detector
simulation as well as the hadronic decay mode of the intermediate $W$
boson.

\begin{table}[hbt]
\begin{center}
\begin{tabular}{|c|c|c|c|} 
\hline $m_r$  & $250$ GeV & $300$ GeV & $350$ GeV\\ 
\hline Signal & $\ a^2_{tc}\times 21$ fb\ & $\ a^2_{tc}\times  15$
fb\ & $\ a^2_{tc}\times  9$ fb\ \\
\hline Background  &  280 fb & 199 fb & 136 fb \\
\hline
\end{tabular}
\end{center}
\vspace{-.4cm}
\caption[]{Signal and background for different Radion masses with
  $\Lambda_r=2$ TeV (and no Higgs-Radion mixing). 
We multiplied by a K-factor of 2.4 for the signal, to account for QCD
  corrections in the Radion production from gluon fusion. \label{tctable}}
\end{table}

In this letter we derived the general flavor structure of the Radion
couplings to bulk fermions, and showed that the same result holds for both bulk and brane
Higgs scenarios. The SM fermion masses and Radion couplings will be
misaligned when the 5D fermion bulk mass parameters are
non-degenerate. This will then lead to FCNC's mediated by the Radion, and
if it is light enough low energy observables such as $\epsilon_K$ will put strong
constraints on the model parameters. 
For a heavier Radion, presumably already disovered through its decay into
Z bosons, one could study the possibility of flavor changing
neutral decays such as $r \to t c$. Although challenging, we estimated that after $300$
fb$^{-1}$ of data an interesting region of the Radion parameter space
could be probed, gaining very valuable information on the flavor
substructure of the whole model.

\acknowledgments
We thank Seung Lee for conversations and specially Kaustubh
Agashe for comments and suggestions.



\begin{thebibliography}{99}


\bibitem{RS1}
  L.~Randall and R.~Sundrum,
  Phys.\ Rev.\ Lett.\  {\bf 83}, 3370 (1999);
  L.~Randall and R.~Sundrum,
  Phys.\ Rev.\ Lett.\  {\bf 83}, 4690 (1999).
  
 \bibitem{a}
  T.~Gherghetta and A.~Pomarol,
  Nucl.\ Phys.\  B {\bf 586}, 141 (2000);
  Y.~Grossman and M.~Neubert,
  Phys.\ Lett.\  B {\bf 474}, 361 (2000);
  K.~Agashe, T.~Okui and R.~Sundrum,
  arXiv:0810.1277 [hep-ph].
  


\bibitem{RSeff}
  K.~Agashe, A.~Delgado, M.~J.~May and R.~Sundrum,
  JHEP {\bf 0308}, 050 (2003).


\bibitem{Burdman}
  G.~Burdman,
  Phys.\ Rev.\  D {\bf 66}, 076003 (2002);
  G.~Burdman,
  Phys.\ Lett.\  B {\bf 590}, 86 (2004);
  S.~J.~Huber,
  Nucl.\ Phys.\  B {\bf 666}, 269 (2003).

\bibitem{Agashe}
 K.~Agashe, G.~Perez and A.~Soni,
  Phys.\ Rev.\  D {\bf 71}, 016002 (2005);
  K.~Agashe, M.~Papucci, G.~Perez and D.~Pirjol,
  [arXiv:hep-ph/0509117].

\bibitem{Weiler}
  C.~Csaki, A.~Falkowski and A.~Weiler,
  JHEP {\bf 0809}, 008 (2008);
  M.~Blanke, A.~J.~Buras, B.~Duling, S.~Gori and A.~Weiler,
  [arXiv:0809.1073 [hep-ph]]; 
  K.~Agashe, A.~Azatov and L.~Zhu,
  [arXiv:0810.1016 [hep-ph]].
  M.~Bauer, S.~Casagrande, L.~Gruender, U.~Haisch and M.~Neubert,
  arXiv:0811.3678 [hep-ph].

\bibitem{GRW}
  G.~F.~Giudice, R.~Rattazzi and J.~D.~Wells,
  Nucl.\ Phys.\  B {\bf 595}, 250 (2001)


\bibitem{CGK}
  C.~Cs\'aki, M.~L.~Graesser and G.~D.~Kribs,
  Phys.\ Rev.\  D {\bf 63}, 065002 (2001);

\bibitem{Korean}
  S.~Bae, P.~Ko, H.~S.~Lee and J.~Lee,
  Phys.\ Lett.\  B {\bf 487}, 299 (2000);
  K.~m.~Cheung,
  Phys.\ Rev.\  D {\bf 63}, 056007 (2001);
  J.~L.~Hewett and T.~G.~Rizzo,
  JHEP {\bf 0308}, 028 (2003);
  D.~Dominici, B.~Grzadkowski, J.~F.~Gunion and M.~Toharia,
  Nucl.\ Phys.\  B {\bf 671}, 243 (2003);
  D.~Dominici, B.~Grzadkowski, J.~F.~Gunion and M.~Toharia,
  Acta Phys.\ Polon.\  B {\bf 33}, 2507 (2002);
  J.~F.~Gunion, M.~Toharia and J.~D.~Wells,
  Phys.\ Lett.\  B {\bf 585}, 295 (2004).


\bibitem{rizzo}
  T.~G.~Rizzo,
  JHEP {\bf 0206}, 056 (2002)

\bibitem{CHL}
  C.~Csaki, J.~Hubisz and S.~J.~Lee,
  Phys.\ Rev.\  D {\bf 76}, 125015 (2007)

\bibitem{mtgamma}
  M.~Toharia,
  [arXiv:0809.5245 [hep-ph]].

\bibitem{Goldberger:1999uk}
  W.~D.~Goldberger and M.~B.~Wise,
  Phys.\ Rev.\ Lett.\  {\bf 83}, 4922 (1999)


\bibitem{bulkhiggs}
  H.~Davoudiasl, B.~Lillie and T.~G.~Rizzo,
  JHEP {\bf 0608}, 042 (2006);
  G.~Cacciapaglia, C.~Csaki, G.~Marandella and J.~Terning,
  JHEP {\bf 0702}, 036 (2007).
  
  
\bibitem{Buras:1998raa}
  A.~J.~Buras,
  [arXiv:hep-ph/9806471].

\bibitem{LRS}
  H.~Davoudiasl, G.~Perez and A.~Soni,
  Phys.\ Lett.\  B {\bf 665}, 67 (2008)


\bibitem{Bona:2007vi}
  M.~Bona {\it et al.}  [UTfit Collaboration],
  JHEP {\bf 0803}, 049 (2008)
  
 \bibitem{Bagger:1997gg}
  J.~A.~Bagger, K.~T.~Matchev and R.~J.~Zhang,
  Phys.\ Lett.\  B {\bf 412}, 77 (1997).

\bibitem{aguilar}
  J.~A.~Aguilar-Saavedra and G.~C.~Branco,
  Phys.\ Lett.\  B {\bf 495}, 347 (2000).


\bibitem{CMS}
  G.~L.~Bayatian {\it et al.}  [CMS Collaboration],
  J.\ Phys.\ G {\bf 34}, 995 (2007), [ATLAS Collaboration],
  CERN-OPEN-2008-020, Geneva, 2008, to appear.


\bibitem{ctburdman}
  G.~Burdman,
  Phys.\ Rev.\ Lett.\  {\bf 83}, 2888 (1999).

  
\bibitem{MTNO}
  N.~Okada and M.~Toharia,
  work in progress.
  
\bibitem{calchep}
  A.~Pukhov,
  [arXiv:hep-ph/0412191].
  
\bibitem{pythia}
  T.~Sjostrand, S.~Mrenna and P.~Skands,
  JHEP {\bf 0605}, 026 (2006)
   


\end{thebibliography}
\end{document}